\documentclass{ws-jbs}
\usepackage[super,compress]{cite}
\usepackage{url}

\begin{document}

\markboth{Szabados, Kerepesi, Bak\'acs}
{MiStImm: a simulation tool to compare CRS immune models with ERS immune model}


\title{MISTIMM: A SIMULATION TOOL TO COMPARE CLASSICAL NONSELF-CENTERED IMMUNE MODELS WITH A NOVEL SELF-CENTERED MODEL}

\author{TAM\'AS SZABADOS}
\address{Department of Mathematics\\ 
Budapest University of Technology and Economics\\ M\H{u}egyetem rkp 3, Budapest, 1521, Hungary\\
\email{szabados@math.bme.hu}
}

\author{CSABA KEREPESI}
\address{Institute for Computer Science and Control\\
Hungarian Academy of Sciences\\
Kende u 13-17, Budapest, 1111, Hungary\\
kerepesi@sztaki.hu}

\author{TIBOR BAK\'ACS}
\address{Alfr\'ed R\'enyi Institute of Mathematics\\
Hungarian Academy of Sciences\\
Re\'altanoda u 13-15, Budapest, 1053, Hungary\\
tiborbakacs@gmail.com}

\maketitle


\begin{abstract}
Our main purpose is to compare classical nonself-centered, two-signal theoretical models of the adaptive immune system with a novel, self-centered, one-signal model developed by our research group. Our model hypothesizes that the immune system of a fetus is capable learning the limited set of self antigens but unable to prepare itself for the unlimited variety of nonself antigens. We have built a computational model that simulates the development of the adaptive immune system. For simplicity, we concentrated on humoral immunity and its major components: T cells, B cells, antibodies, interleukins, non-immune self cells, and foreign antigens. Our model is a microscopic one, similar to the interacting particle models of statistical physics and agent-based models in immunology. Furthermore, our model is stochastic: events are considered random and modeled by a continuous time, finite state Markov process, that is, they are controlled by finitely many independent exponential clocks.

The simulation begins after conception, develops the immune system from scratch and learns the set of self antigens. The simulation ends several months after birth when a more-or-less stationary state of the immune system has been established. We investigate how the immune system can recognize and fight against a primary infection. We also investigate under what conditions can an immune memory be created that results in a more effective immune response to a repeated infection. The simulations show that our self-centered model is realistic. Moreover, in case of a primary adaptive immune reaction, it can destroy infections more efficiently than a classical nonself-centered model.

Predictions of our theoretical model were clinically supported by autoimmune-related adverse events in high-dose immune checkpoint inhibitor immunotherapy trials and also by safe and successful low-dose immune checkpoint inhibitor combination treatment of heavily pretreated stage IV cancer patients who had exhausted all conventional treatments.
The MiStImm simulation tool and source codes are available at the address https://github.com/kerepesi/MiStImm.
\end{abstract}




\section{Introduction}

\subsection{Motivation of our work} \label{ssec:History}

The vast majority of published papers still portray the immune system with many idealizations that neglect important epidemiologic observations and experimental data at the expense of biological commonsense, see Section 1.1 in Ref.~\refcite{SB11}. For example, the accepted dogma still claims that the evolution of the immune system was driven by pathogens and a clonally based immune system is capable of efficiently fighting primary bacterial or viral infections. Several observations, however, cannot be reconciled with such assumptions.

US (Table 3 in Ref.~\refcite{CM01}) and Hungarian~\cite{MSE1896, Ker16} records from 1900 and 1896, respectively, before the dramatic medical advances, show 32\% and 27\% deaths attributable to infections, whereas only 5\% and 2\% due to cancer. The situation is similar even nowadays in the case of low income countries~\cite{WHO12}. These data demonstrate that the immune system is far from being infallible against pathogens. In contrast, the low cancer incidence can be interpreted to mean that the immune system primarily evolved to ``maintain individual integrity in the midst of chaotic communal living''~\cite{Rin04} and just sequentially to cope with pathogens.

Considering the historical low death rate from cancers versus the high death rate from infections, furthermore, the very slow proliferation of cancer cells versus the explosive replication speed of pathogens, we argued for a \emph{self-centered model} as the explanation for T cell activation versus tolerance; see the long history of this view of the immune system in Ref.~\refcite{San15}. Based on information theoretical principles and the law of parsimony we suggested that the ability of the immune system to recognize all kinds of self antigens is sufficient to attack any nonself antigen~\cite{BMSVT01,BSVT01,BMSVST07,SB11}. In order to discriminate self and nonself, a relatively large fraction of T lymphocytes -- the set of regulatory T cells (Treg cells) -- should primarily recognize the much smaller and always available set of self antigens, rather than the practically unlimited and for the immune system only partially known nonself antigen universe. In our model, the role of regulatory T Cells (Foxp3+ Tregs) seems to be the closest analogy to the role of homeostatic T cells.

Predictions of our theoretical model were supported by numerous clinical trial observations. Immunotherapy has become a very promising approach to treat cancer in the last few years. However, the developers of the inhibitory anti-CTLA-4 antibody started with the premise that a CTLA-4 (cytotoxic T lymphocyte-associated antigen 4) blockade would selectively target T cells involved in the anti-tumor immune response~\cite{CCSA12}. Although the anti-CTLA-4 antibody improved survival in a minority of metastatic melanoma patients, the vast majority suffered autoimmune-related adverse events (irAEs)~\cite{BKBTS15}.

While the conventional nonself-centered, two-signal T cell activation models are unable to explain the widespread and dose-dependent irAEs, our self-centered, one-signal T cell activation theory can~\cite{BMSVST07}. The reason for this that tolerance mechanisms of the nonself-centered, two-signal models eliminate self-reactive immune cells to ensure that signal one can only originate from a foreign/mutated antigen. Immune cells, however, require cognate receptor engagement with ubiquitous self antigens in their `flight for survival'~\cite{FR00}. Our model, therefore, predicted that a large ratio of T cells should be temporarily activated by self antigens thus expressing CTLA-4 receptors that can be engaged by anti-CTLA-4 antibodies. It is consistent with the \emph{immunological homunculus} concept of Irun Cohen, who suggested that the immune system continuously responds to self~\cite{Coh92}.

Nothwithstanding, prolonged overstimulation of T cells by antibodies that target their negative regulators (immune checkpoint, IC) such as CTLA-4 and the programmed cell death protein 1 pathway (PD-1/PD-L1) led to a breakthrough in the treatment of a variety of malignancies. Although three generations of IC immunotherapy have been developed since Ref.~\refcite{All15,Hoo16}, the safety of IC blockade is still an unresolved, timely and sensitive issue in the context of advanced cancer patients.

Based on our self-centered, one-signal theory, we have addressed the controversy regarding the safety--efficacy issue in certain immunotherapy trials and argued that the price we pay for reversing immunosuppression in cancer by a prolonged immune checkpoint blockade is the generation of uncontrolled T-cell activation~\cite{BMM12,BMSM12,SMB13,BM15}.
In fact, we predicted that harnessing the unleashed autoimmune power of T cells by low dose IC blockade could be rewarding to defeat cancer. Using our prediction, Ref.~\refcite{KMSBB16} have developed just such a promising combination therapy, which was safely and successfully administered to heavily pretreated stage IV cancer patients who had exhausted all conventional treatments.

\subsection{Theoretical and computational models} \label{ssec:Theo}

In a wide class of theoretical models~\cite{Jwbook,Paulbook} even a primary immune reaction depends on the recognition of nonself antigens by T and B cell receptors, so the theory is \emph{nonself-centered}. The role of self in those models is that the great majority of autoreactive T and B cell clones are selected and purged from the immune system~\cite{FVGB2004}. For brevity, such theoretical models will be called \emph{Conventional Role of Self models} \emph{(CRS models)} in the sequel.

On the other hand, a smaller class of theoretical models is based on the assumption that recognizing and preserving self is the primary task of the immune system; these are the \emph{self-centered} models~\cite{San15}. Our group's theoretical model belongs to the class of self-centered models. It hypothesizes that the immune system of a fetus can primarily learn what self is but is unable to prepare itself for the huge, unknown variety of nonself. Consequently, a \emph{primary} reaction against a nonself antigen is possible just by recognizing that the new antigen is not self. The assumed intrauterine learning process results in a repertoire of regulatory T cells (Tregs) that plays a fundamental role: \emph{the set of Tregs keeps the immune image of the set of self antigens during whole life} and so -- beside defending \emph{self} from autoimmune reactions, as in conventional models -- \emph{directs immune reactions against nonself}. Our theory will be called \emph{Enhanced Role of Self model} \emph{(ERS model)} from now on.

Similar (but not identical) to our model is the mathematical model of T cell mediated suppression of Ref.~\refcite{LLC03}, where tolerance is also based on ubiquitous and constitutive self-antigens, which select and sustain clones of specific regulatory (R) cells, and which are similar to our Treg cells. In their model R cell populations represent typically between 30\% and 95\% of the total T cells in the periphery. It is an important difference to the widely accepted view in which conventional regulatory CD4+CD25+ T cells (Treg) usually make up only about 5\%--10\% of CD4+ T cells~\cite{CKL2016}. R cells perform their function through linked recognition of the APCs (antigen presenting cells). Also in their model, immune responses to foreign antigens are achieved by displacing the self-antigens from the APCs, leading to a loss of R cells if the foreign antigen introduction entails a sharp increase in the number of foreign antigen carrying APCs.

Further, our intention was to create a computational model as well to show that the ERS conceptual model is able to work in silico as is expected from the immune system. Moreover, we wanted to show that the ERS model performs better than CRS models in silico.

Table 1 in Ref.~\refcite{GMNF11} broadly classifies computational models in immunology into four groups: (1) individual particle-based stochastic, (2) particle number stochastic, (3) concentration-based spatial non-stochastic, (4) concentration-based non-spatial non-stochastic (see 330 references therein). A very broad class of computational models uses ordinary differential equations and belongs to (3) or (4). Another wide model classes are the cellular automata and agent-based models, belonging to (1) or (2). Our computational model is in part individual particle based and in part particle number stochastic, (1) and (2) combined. Essentially, we employed the ideas of agent based models, though we used exclusively our own software.

A great advantage of such a model is that it can easily incorporate the most important types of cells and molecules together with their essential features and events that play important roles in immune reactions. In such a model events -- for example interactions of components -- occur at random. Also, such a model is typically microscopic in space and limited to a small variety of cells and molecules.

A stochastic model fits well with the affinity maturation of B lymphocytes in which random events are perhaps the most characteristic. It is also suitable to model the development of the regulatory T cell population and the random selection of specific T cell clones. A major advantage of this approach is that it permits studying random variations in the immune process.

To simplify things, we chose the humoral adaptive immune system as the first modeling objective, since the humoral phase (blood or lymph) may be considered spatially homogeneous; thus a microscopic spatial volume may represent the whole phase well. A major advantage of this approach is that it is not necessary to describe the actual spatial positions and spatial motions in the model. Instead, model components randomly choose one of the other components as interaction partners, because any components are close enough to become engaged in an interaction.

In sum, the novelty of the present paper is partly our ERS model which is a specific self-centered conceptual immune model and partly the MiStImm computational model that we have developed to compare different theoretical models in silico. The main research question of our work is to decide in silico if the ERS model is feasible and it is able to fight against infections; moreover, whether it can fight more efficiently than CRS models. 

\subsection{Some related conceptual and computational models} \label{ssec:rel_ref}

Important precursors to our work, using self-centered stance, were several models by I.R. Cohen and coworkers~\cite{HSC2003,GP2001,CHS2004,EHC2007,VHCE2011}.

To our best knowledge, the first experiments with a detailed agent-based model (IMMSIM) of immune system were~\cite{CS92,SC92,CS96}. Their goal was to capture the dynamics of the immune system and to perform experiments in silico. Later they studied the thymus, the regulation of positive and negative selection, and the dynamics of the production of the TCR repertoire in the thymus~\cite{MSSC95}. Computational models mainly based on the idea developed by Celada and Seiden have been also used in cancer immunology; a review is in Ref. \refcite{PPPCM12}.

A closely related agent-based model, the \textsc{C-ImmSim} package has been developed and investigated in Ref.~\refcite{CBS97}. Later it was modified by Rapin et al; an excellent recent description of their work can be found in Ref.~\refcite{RLBC10}. Their model represents pathogens, as well as lymphocytes receptors, by means of their amino acid sequences and makes use of bioinformatics methods for T and B cell epitope prediction. This is a key step for their simulation of the immune response, because it determines immunogenicity. The related book~\cite{CC15} can be used as a practical guide to implement a computational model with which one can study a specific disease.

The Basic Immune Simulator (BIS)~\cite{FAO07} is also an agent-based computing model to study the interactions between innate and adaptive immunity. The BIS was created using the Recursive Porus Agent Simulation Toolkit (RepastJ) library, an open-source software library that is available online~\cite{NCV2006}.

Ref.~\refcite{KCHSN06} have developed SIMISYS, which is also a cellular automata model of the human immune system. It uses tens of thousands of cells and innate and adaptive components of the immune system. In particular, the model contains macrophages, dendritic cells, neutrophils, natural killer cells, B cells, T helper cells, complement proteins, and pathogenic bacteria.

Ref.~\refcite{CBR13} investigates a hypothesis about B cell hypermutation and affinity maturation using both individual particle based stochastic and concentration-based non-spatial non-stochastic, ordinary differential equation models.

Finally, we mention an important recent book about immune system modelling~\cite{MDDMMZ15}. In particular, a B cell model is developed in Ref.~\refcite{SWB15}; the model has partly similar ideas as our B cell model, but differs from ours in the representation of ligands. Ligands in their model are encoded by bit strings and their distances are measured by the number of mismatches (Hamming distance). It can be mentioned that this kind of representation of ligands has appeared in many earlier models like IMMSIM and C-ImmSim as well.


\section{The ERS theoretical model}

As was mentioned, our ERS model belongs to the class of self-centered models. Here we describe the major aspects of our model. 

\subsection{A single T cell cannot discriminate self and nonself, only a wide Treg repertoire can} \label{ssec:Treg}

Shapes of self and nonself entities are intricately interwoven sets; in the language of the shape space model, the subsets of points representing self and nonself are complexly interlaced and cannot be separated by a nice smooth mathematical curve. Therefore the complexity of the antigen universe exceeds the capacity of an individual T cell. The ``knowledge'' of each specific T cell is reflected by the shape of its TCR. An individual T cell therefore is able to recognize only a set of complementary or near complementary MHC-peptide molecule. In the present paper T cells with nearly complementary TCR to self-MHC-peptide complexes are designated as \emph{regulatory T cells, Treg cells}~\cite{Wil13}.

In particular, the complete repertoire of Treg cells is able to reflect the whole set of self antigens (See Ref.~\refcite{SB11} and Fig. 1 and video animation in Ref.~\refcite{BMSVST07}). The repertoire of Tregs is first created in the thymus of the fetus by negative and positive selection and it constitutes the basis for self--nonself discrimination. Any self-MHC-peptide complex that is able to attach to a Treg with intermediate affinity can be classified as \emph{self}; any other MHC-peptide complexes -- that has weak affinity to each Treg but may have strong affinity to one of the T cells -- can be classified as \emph{nonself}. Thus the Treg repertoire -- like the conductor of an orchestra -- controls other elements of the adaptive immune system. This does not exclude the possibility that Tregs -- like players of an orchestra -- may take part in immune reactions similarly to other conventional T cells as well. See further details in Ref.~\refcite{SB11}. After birth, development of infection specific T cell and B cell clones are under Treg control.

Treg cells turn off antibody production and suppress the immune response. The details of Treg cells functioning are still debated~\cite{Cor2009,HTV2013}. For example, it is not clear whether Treg cells can directly suppress B cells or whether they must suppress Th cells in order to suppress B cells. Similarly as in Ref.~\refcite{Ann2017}, we model the direct suppression of B cells, which has been suggested in a number of recent studies, see e.g. Ref~\refcite{WZ2013,GGL2012}. 

\subsection{Different T cell -- B cell interactions}\label{ssec:Interactions}

\begin{figure}[ht]
     \centering
     \includegraphics[width=0.8\textwidth]{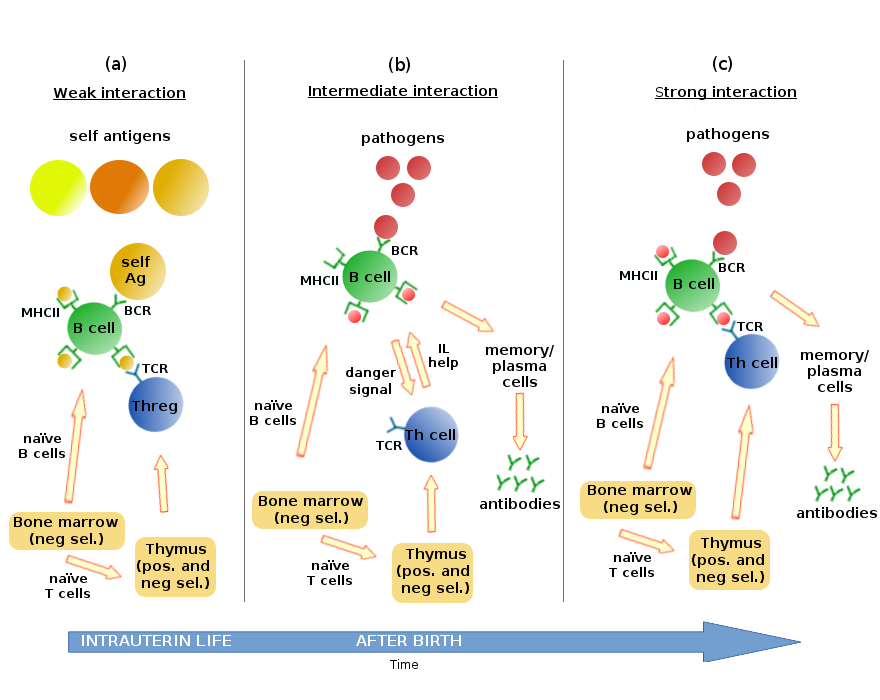}
    \caption{{\bf Humoral adaptive immune response.} 
The ERS model is described by (a), (b) and (c), while CRS models are described by (c) alone. 
In the ERS model, weak affinity interaction (a) begins in intrauterin life and keeps the immune image of self during whole life. Also in the ERS model, intermediate affinity interaction (b) is the initial phase of a primary infection. Strong affinity interaction (specific immune reaction) (c) appears in both the ERS and CRS models and usually needs several days to efficiently start. Signal strength (irrespective whether it comes via one or two receptors) determines the outcome of B cell activation and/or clonal expansion. \emph{Weak affinity interaction (a)} is sufficient just for homeostasis; low affinity BCR binds self-antigens and presents self-peptides in MHCII to regulatory T helper (Threg) cells; this ensures B and Threg cell survival. \emph{Intermediate affinity interaction (b)} is required for eradicating primary infections; some B cells that have higher BCR affinity for the antigens of the pathogen than that of the host will capture  pathogens with intermediate affinity and present pathogen derived foreign peptides in MHCII. The foreign peptides inhibit binding of Threg cells to these B cells for a critical time period, then the latter will secrete soluble danger signals. Danger signals activate local Th cells, which in turn, release interleukins that fuel local T cell activation, both helper and cytotoxic T cells. Eventually a local cytokine storm is generated. This way a non-specific, local polyclonal B and T cell activation is induced, which is the defense mechanism against primary infections in the ERS model. Clonal expansion requires affinity maturation, which results in a several magnitude increase of BCR affinity, typically over a time of one week. Random mutations cause the production of B cells with a broad range of affinities for their antigen. B cells with unfavorable mutations will not get sufficiently activated by the antigen and will die, while those with improved affinity will be stimulated to clone themselves. This leads to an effective affinity-dependent selection process. 
\emph{Strong affinity interaction (c)} in the ERS and CRS models, in contrast, is supervised and supported by pathogen peptide specific Th cells, which require direct contact via TCR to the MHCII of the expanding B cell clone. This process is significantly slower than (b).} 
     \label{fig:immune_response}
\end{figure}

As in our current computational model T cell -- B cell interactions are basic, here we describe three different types of it. Each of the three types fulfills an important role in the ERS model (Fig.~\ref{fig:immune_response}). Typical CRS models can be described by the third type of interactions alone.

In a healthy individual during intrauterine life, randomly produced moderately self-reactive B cell clones are confronted with an overwhelming quantity of soluble self antigens. Those B cells that can attach with intermediate affinity to any of these self antigens via their B cell receptors (BCRs) will present self peptides in their surface major histocompatibility complex II (MHCII) molecules to regulatory T helper cells (Thregs). This ensures B cell and Threg cell survival, respectively, but it is insufficient to trigger extensive clonally based B cell expansion required for specific immunity or autoimmunity. It will be called \emph{weak affinity interaction and division} from now on. Thus the positively selected Threg cells are critical parts of the homeostatic control in our model, so that Threg clones exist for practically all kinds of self-MHCII -- self-peptide complexes presented by any of the B cells. After birth, this process maintains \emph{an immune image of soluble self which can control self--nonself discrimination}.

During a \emph{primary infection} a new antigen appears in the blood. B cells with appropriate affinity for the new antigen, engulf new antigens and present its foreign peptides on their surface MHCII proteins. Since in our model foreign peptides transiently inhibit the complementary TCR-MHC interactions, such perturbation creates steric hindrance that obstructs the docking of positively selected Thregs. Disruption of such contact between an MHCII and Thregs \emph{for a critical period of time} results in an emergency and activates the corresponding B cell. In order to reestablish contact, foreign peptide presenting B cells will secrete \emph{chemotactic danger signals} \emph{(``smoking gun'')} attracting Th cells to this region. The B7-1 and B7-2 ligands of B cells will activate most CD28 receptors of the bystander helper T cells. This initiates a non-specific, polyclonal activation in local Th lymphocytes via the CD28 receptor alone~\cite{MSB07} such that a local cytokine storm is generated in Th cells triggering B cells to clonal expansion, hypermutation, and eventually they may develop into specific antibody producing plasma cells. This will be called \emph{intermediate affinity interaction and division} from now on. The resulting inner state of the affected Th and B cells will be called \emph{``activated''} state. Since affinity  maturation is driven by the fast increasing local concentration of pathogen antigens (e.g.\ hepatitis virus), the probability of clonal autoimmunity is very low but possible. 

The default mode of our model is that a random peptide decreases complementarity between a na\"{\i}ve TCR and the MHC. However, following the initial polyclonal activation phase, there is always a possibility that rare T cell and B cell clones with higher affinity may well recognize foreign antigens, particularly when a significant fraction of host cells is infected and viral load is high (for example in hepatitis, see in Ref.~\refcite{TOC01}). Such higher affinity interactions would then drive clonal (e.g. HCV specific) T cell proliferation, activation, lysis of infected cells, as described by the conventional two-signal models. Having cleared the infection, specific T cells could eventually become an expanded memory type T cell clone, while B cells could differentiate into infection specific antibody producing plasma cells or memory B cells. It is thought that acquisition of memory T cell function is an irreversible differentiation event. Unlike regulatory T cells, such population does not require self-peptide--MHC complexes for maintenance. Nevertheless, sustaining the functional phenotype of T memory cells requires active signaling via CD27~\cite{ACG09}. Specific T and B cell activation, proliferation and lysis of infected cells, therefore, obey the rules of the conventional two-signal model. Clearly, this process may require several days in general. It will be called \emph{strong interaction and division} in the sequel. The resulting inner state of the affected Th and B cells will be called \emph{``strongly activated''} state.

\section{Description of the MiStImm computational model}

We made an effort to realize the above-mentioned ERS conceptual model in a computational model as accurately as possible. The ERS theoretical model implies that the immune system is a complex system; consequently, our computational model has to be a complex model as well. We tried to stick to experimental facts and pure logic as much as possible. Notwithstanding, we have to admit that several authors are sceptic about such complex models of immune dynamics, see e.g. Ref.~\refcite{OF14}. On the other hand, Ref.~\refcite{EC02} convincingly argue that the immune system is a complex system, thus a ``minimal model'' like Ref.~\refcite{LC02} cannot describe the behavior of immune system correctly.

We call our in silico model \emph{Mi}croscopic \emph{St}ochastic \emph{Imm}une model or briefly \emph{MiStImm} model. It is a further developed version of our 1994-98 B cell model~\cite{STVB98}. Our software is a C program and it was written in the spirit of the agent-based models.

\subsection{Mathematical model} \label{ssec:Basics}

Mathematically, the interactions of the components and other events in the model are described by a continuous time, finite state, time-homogeneous Markov process, see e.g. Ref.~\refcite{Lam12}. A Markov process is a memoryless stochastic process: if we specify the present state of the system, then we may forget about its history when we want to investigate its behavior in the future.

More precisely, if the possible states of the system are denoted by the natural numbers $1, 2, \dots, M$, and $X_t$ is the random state of the process at time $t \ge 0$, then the process is described by the transition probabilities
\[
P_{i,j}(t) = \mathbb{P}(X_{t+s} = j \mid X_s = i) \qquad (i,j = 1, \dots, M; \quad s,t \ge 0).
\]
Let $\mathbf{P}(t) = [P_{i,j}(t)]_{i,j =1}^M$ and suppose that $\mathbf{P}(0) = \mathbf{I}$ and $\lim_{t \to 0^+} \mathbf{P}(t) = \mathbf{I}$, where $\mathbf{I}$ is the identity matrix. Then it is well-known that
\[
\mathbf{P}(t) = e^{\mathbf{Q} t} = \mathbf{I} + \mathbf{Q} t + \frac12 \mathbf{Q}^2 t^2 + \cdots, \qquad \mathbf{Q} = [q_{i,j}]_{i,j = 1}^M,
\]
where $\mathbf{Q}$ is the infinitesimal generator of the Markov process. Thus
\begin{equation}\label{eq:Pij}
P_{i,j}(t) = \delta_{i,j} + q_{i,j} \, t + o(t) \quad \text{as} \quad t \to 0^+ ,
\end{equation}
where $\delta_{i,j} = 1$ if $i=j$ and $0$ otherwise, and $o(t)/t \to 0$. It means that the probability of a transition from state $i$ to a state $j \ne i$ is determined by the rate $q_{i,j} \ge 0$; $q_{i,i} = - \sum_{j \ne i} q_{i,j}$.

Let us recall that when one has a Markov transition probability
\[
P_{i,i+1}(t) = q t + o(t), \quad P_{i,i}(t) = 1 - q t - o(t) \quad \text{as} \quad t \to 0^+ ,
\]
then dividing the time interval $[0,t]$ into $n$ equal subintervals, it follows for the corresponding Markov process $Y_t$ when $Y_0=0$ that
\[
\mathbb{P}(Y_t = k) = \lim_{n \to \infty} \binom{n}{k} \left(\frac{qt}{n} + o\left(\frac{t}{n}\right)\right)^k \left(1 - \frac{qt}{n} - o\left(\frac{t}{n}\right)\right)^{n-k}
\]

\[= e^{-q t} \frac{(q t)^k}{k!}  \quad (t \ge 0, k = 0, 1, 2, \dots).
\]

Thus $Y_t$ is a Poisson process, and so the \emph{holding time} $T := \inf\{t \ge 0 : Y_t \ne 0\}$ is \emph{exponentially distributed}:
\[
\mathbb{P}(T \ge t) = e^{-q t} \qquad (t \ge 0) .
\]

Hence it follows from (\ref{eq:Pij}), that if $X_s=i$, the holding time $T_i := \inf\{t \ge 0 : X_{s+t} \ne i\}$ is also exponentially distributed:
\begin{equation}\label{eq:Qi}
\mathbb{P}(T_i \ge t) = e^{-Q_i t} \quad (t \ge 0) , \quad Q_i := \sum_{j \ne i} q_{i,j} = - q_{i,i}.
\end{equation}

Thus one can realize the Markov process $(X_t)_{t \ge 0}$ by assigning to any potential random event an independent exponential clock with rate $q_{i,j}$ $(j \ne i)$, supposing that the present state of the system is $X_s = i$. When the first clock rings, say, the $j$th one, the corresponding event, that is, the change from state $i$ to $j$, occurs with rate $q_{i,j}$.

The simulation uses the well-known fact that when there are independent exponential clocks with rates $q_{i,j}$ $(j \ne i)$, then the fastest event has also exponential clock with rate $Q_i := \sum_{j \ne i} q_{i,j}$, see (\ref{eq:Qi}). So at any step, it is enough to generate a single exponential random number with rate $Q_i$. Also, the probability that the event $j$ has occurred, is equal to $q_{i,j}/Q_i$ $(j \ne i)$, whose sum is $1$. Thus one generates a uniform random number in $[0, 1)$, and its value determines which one of the concurrent events has occurred.

Our model has finitely many components at any time $t$: helper T cells (regulatory Th cells and potential infection specific Th cells), B cells, antibodies, interleukins, non-immune self cells, and foreign antigens. Presently, other than helper type T cells or other antigen presenting cells besides B cells are not represented in our computational model. Each component has a number of characteristics (parameters) and certain attached random events or processes of events that may occur at random. A potential event can be, for example, a division of a cell or an interaction of a component with a randomly chosen partner. The occurrence of such an event may cause several changes in the model, like births, deaths, and updates of parameters. Because of the births and deaths, our mathematical model is somewhat more general than the simple Markov model described above: the size $M$ of the system changes with time in general. However, it does not cause much difference. We set the model parameters so that explosion does not occur and hence the number of potential events remain finite for all $t$. At any step, one has to establish the number of independent exponential clocks $M(s)$, and determine their actual rates $q_{i,j}=1/\tau_{i,j}$ $(1 \le j \le M(s), j \ne i)$, where $\tau_{i,j}$ is the \emph{mean holding time} of the $j$th component. Then the simulation starts again with the new settings.

\subsection{Basics} \label{ssec:Tools}

\paragraph*{Peptide lattice} \label{sssec:peptide}
Our computational model takes a microscopic volume of the humoral phase and also a microscopically small part of the shape space universe. Shape space models were used by Perelson, Segel and their colleagues since the 1970's~\cite{PO79,SP88} and also in the Celada--Seiden models mentioned above. To explain what we mean by shape space here, assume that the shape of a T cell receptor (TCR) can be represented by a point in a large set of a Euclidean space. Theoretical considerations compared with experimental data led to the conclusion~\cite{PO79} that the dimension of this shape space, i.e. the number of parameters essential in describing a binding, is not too large, probably around five.

\begin{figure}[ht!]
\centering
\includegraphics[scale=1]{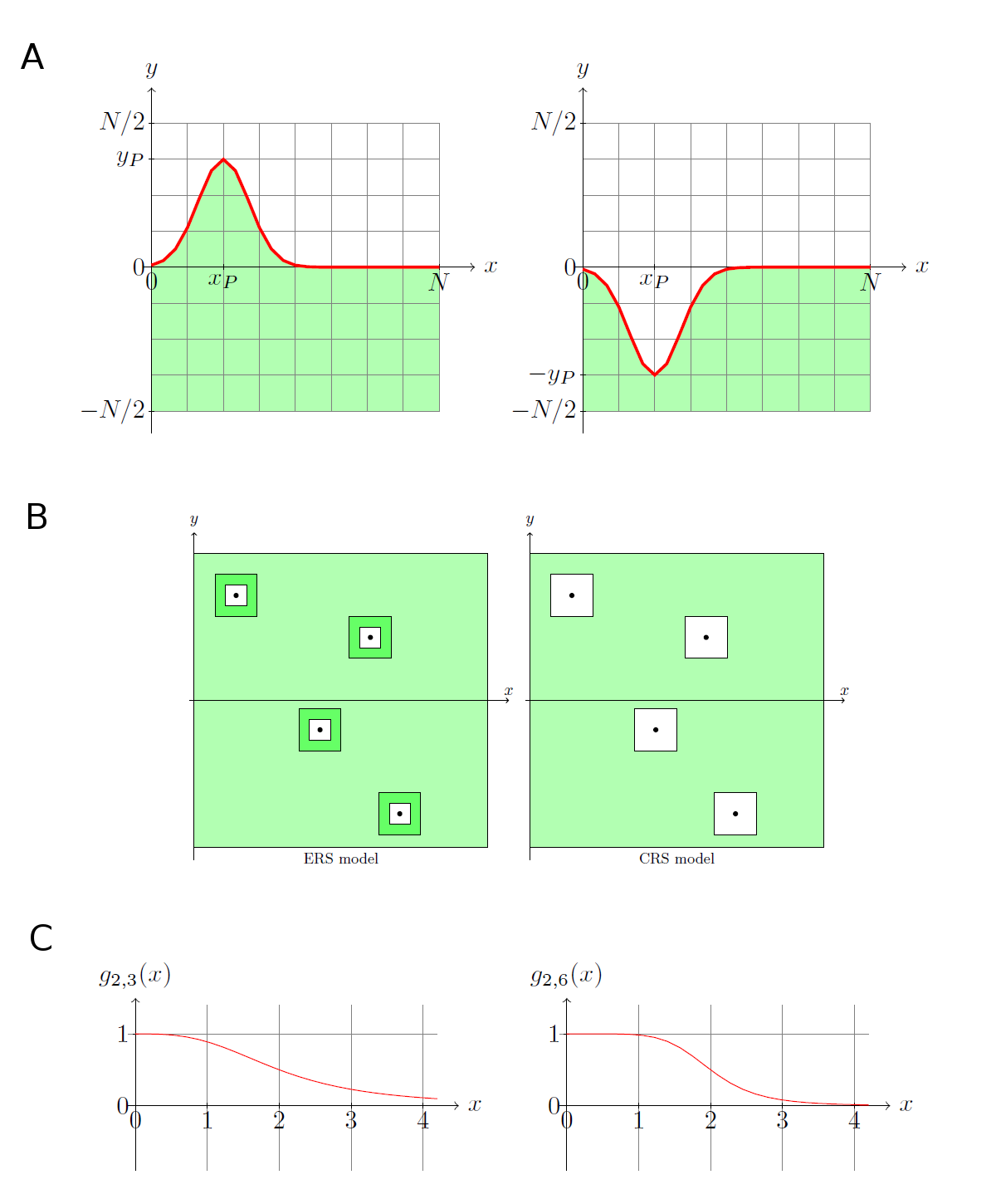}
\caption{{\bf Figures of the mathematical model.} ({\bf A}) Shape space. Two simplified complementary shapes characterized by the points $(x_P,y_P)$ and $(x_P,-y_P)$, respectively, in the peptide lattice. ({\bf B}) Simplified graphical representation of the difference between the ERS and the CRS models. Dark green: area allocated to regulatory T cells; light green: area for potential infection specific T cells. ({\bf C}) Two examples of a logistic function.}
\label{fig:math_model}
\end{figure}

The microscopically small part of the shape space that we consider in our model is a small discrete $N \times N$ planar grid in the shape space (e.g. $N=1000$). The $x \in \{0, 1, \dots , N \}$ coordinate of a shape point may represent a ``horizontal'' coordinate of the main part of the binding profile of a TCR or an MHC+peptide complex, while the $y \in \{-N/2, \dots, N/2 \}$ coordinate may represent the ``vertical'' coordinate of the main part of the binding profile. A positive coordinate represents ``convexity'', while a negative coordinate represents ``concavity''. Fig.~\ref{fig:math_model}A shows our underlying idea for the shape of a peptide characterized by a single point $(x_P,y_P)$. Needless to say that our model of shapes is a much simplified one, but is still suitable to represent essential binding properties of antigens. We call the above finite square grid the peptide lattice in the sequel.

\paragraph*{Antigen lattice}
Shape of a B cell receptor (BCR) or shape of an antigen is similarly represented by a point of an antigen lattice in the model. Here again the $x \in \{0, 1, \dots , N \}$ coordinate of a shape point may represent a ``horizontal'' coordinate of the main part of the binding profile of the BCR or antigen, while the $y \in \{-N/2, \dots, N/2 \}$ coordinate may represent the ``vertical'' coordinate of the main part of the binding profile; a positive coordinate representing ``convexity'', while a negative coordinate representing ``concavity'', see Fig.~\ref{fig:math_model}A.

For simplicity, to each antigen $(x_A, y_A)$ in the antigen lattice we assign exactly one peptide $(x_P, y_P)$
in the peptide lattice. To make identification of an antigen and its corresponding peptide easier, we will use the convention that $x_A = x_P$ and $y_A = y_P$.

\paragraph*{Complementarity} \label{sssec:compl}
Complementarity plays a basic role in binding. The perfect fit between a TCR and an MHC+peptide complex means in the model that the shape $(x_T,y_T)$  of the TCR and the shape $(x_P, y_P)$ of the MHC+peptide satisfy the equalities $x_T=x_P$ and $y_T = - y_P$, see Fig.~\ref{fig:math_model}A. In the model we introduce \emph{a metric} or \emph{distance} $d$ to measure the degree of similarity of two shapes $z_1 := (x_1, y_1)$ and $z_2:= (x_2, y_2)$:
\[
d(z_1, z_2) := \max \{|x_2 - x_1|, |y_2 - y_1| \} .
\]
A TCR $z_T := (x_T,y_T)$ and an MHC+peptide $z_P := (x_P, y_P)$ are \emph{nearly complementary} in our model if the distance between $z_T$ and $\overline{z_P} := (x_P, -y_P)$ is small enough. Similar is the representation of the complementarity between BCRs and antigens in the model.

Only complementary or nearly complementary shaped ligands and receptors can bind. The dots in Fig.~\ref{fig:math_model}B represent TCRs that are exactly complementary to some self MHC+self-peptide complex. The areas shaded in darker green are called the \emph{characteristic rings of self-peptides}. They represent the set of shapes that are allocated to possible \emph{regulatory T cells} after negative and positive selection in the ERS model, see below. The areas denoted by lighter green correspond to possible shapes of classical, \emph{potentially infection (or mutation) specific T cells}, while white areas are representing self-reactive T cells that are prohibited for T cells in the two respective models. Observe that in the ERS model, moderately self-reactive T cells are present after negative and positive selection. In fact, they constitute the most important class of T cells that decide self--nonself discrimination. On the other hand, such moderately self-reactive T cells are negatively selected out in CRS models.

We mention that with the above metric, ``circles'' are in fact squares in our shape space model.

\paragraph*{A logistic function}
In biology it is typical that when the size of a certain cell population gets larger the per capita birth rate in the population decreases. Thus the size of a population first increases fast, later it slows down, and at the end it gets relatively stable. So to control birth rates and other quantities we use a class of logistic functions, previously applied by many other authors (see e.g. Ref.~\refcite{KUT85,BP91}):
\begin{equation}\label{eq:logisticf}
g_{\theta, \eta}(x) := \frac{\theta^{\eta}}{\theta^{\eta} + x^{\eta}} = \left(1 + \left(\frac{x}{\theta}\right)^{\eta}\right)^{-1} \qquad (x \ge 0; \theta > 0, \eta > 0).
\end{equation}
This formula describes a decreasing function which is equal to 1 for $x = 0$, $1/2$ for the threshold value $x = \theta$, and goes to 0 as $x \to \infty$, see Fig.~\ref{fig:math_model}C. Its parameters $\theta$ and $\eta$ are set from case to case. We set the model parameters so that explosion does not occur. In fact, the number of components should always remain in the biologically feasible domain.

\subsection{Self cells} \label{ssec:Self}

At time zero, there is a number (say, 3) of different types of non-immune self cells (briefly: \emph{self cells}), each with a given initial population size (e.g. 150). A certain type of self cells is represented by its position $(x_S, y_S)$ in the antigen lattice and its peptide $(x_P,y_P)$ in the peptide lattice. Specifically, there is a population of \emph{bone marrow cells}, handled separately from other self cells, with a given initial population size.

Each type of self cells comes with a birth process with a given initial rate (that is, with a given initial average waiting time $\tau_{s0}$ between divisions). If the size of the population of a specific self cell at a certain time $t$ is $s=s(t)$, then the conditional expected waiting between two divisions in this population is
\begin{equation}\label{eq:selfbirth}
\tau_s = \frac{\tau_{s0}}{s \, g_{\theta,\eta}(s)} = \frac{\tau_{s0}}{s} \left(1 + \left(\frac{s}{\theta}\right)^{\eta}\right), \quad (\eta > 1).
\end{equation}
Formula (\ref{eq:selfbirth}) indicates that when the number $s$ of a type of self cells becomes significantly larger than its threshold value $\theta$ its division rate gets close to zero. For the sake of simplicity, the natural death process of self cells is not represented in the model, so, more accurately, (\ref{eq:selfbirth}) should be called the effective growth model of self cells. 

We assume that the concentration of each type of self antigens in the humoral phase is directly proportional to the number of self cells carrying this antigen.

The case of bone marrow cells is special because it comes not only with a birth rate, but, with given rates, bone marrow cells also produce na\"{\i}ve B cells and Th cells. Na\"{i}ve B and Th cells have randomly determined BCR and TCR shapes that are uniformly distributed on the antigen and peptide lattices, respectively.

\subsection{Danger signals and interleukins} \label{ssec:Inter}

We use the symbolic names ``danger signals and interleukins'' in this paper, without specifying the exact type of these molecules, similarly to Fig.~3 of Ref.~\refcite{ECDM15}. These types of soluble molecules have roles only in intermediate interactions and divisions in the ERS model. Since conventional immune reactions correspond to the ones that we call strong interactions and divisions, these types of molecules do not appear when simulating CRS models.

\emph{Danger signals (soluble molecules)} are emitted by B lymphocytes following disruption of homeostatic complementary interaction of B cells and Threg cells. This process initiates an action process and also a death process of these molecules. Each danger molecule randomly chooses a Th cell. This is a signal for the Th cell to start intermediate type division and to secrete interleukins. Note that this danger signal is not the same as in Ref.~\refcite{M2002} because our danger signals are emitted when the system detects any kind of nonself and not only a dangerous one. 

\emph{Interleukins} are emitted by Th lymphocytes. This process initiates an action process and also a death process of these interleukins. Each interleukin molecule randomly chooses a B cell. This is a signal for a B cell that has lost complementary Threg cell control to start cell division of intermediate kind.

\subsection{Th cells} \label{ssec:Th}

While different types of non-immune self cells and foreign cells (pathogens) are treated as populations, B and Th cells are handled individually in the model. Pre-Th cells are born in the bone marrow. The birth of a pre-Th cell initiates its own (natural) death event, a Th cell action process and a Th cell activation control process.

\paragraph*{Th cell recognition region}
Each Th cell has a recognition region in the peptide lattice. If a TCR is described by the point $(x_T, y_T)$, then the corresponding recognition region is a square with center $(x_T, -y_T)$ and radius $r_T$. The radius of the TCR is a constant, there is no hypermutation or affinity maturation for Th cells. The recognition region describes the potential shapes of antigens with which a TCR can bind: the smaller the distance between a peptide $(x_P, y_P)$ located on an MHCII and the center $(x_T, -y_T)$ of the recognition region of the TCR, the better the fit.

\paragraph*{Thymus}
To each pre-Th cell there is assigned a random event that places it into the thymus. Here the Th cell goes under a negative and a positive selection process. \emph{Negative selection} kills pre-Th cells that are \emph{closer to one of the self-peptides} than a minimum radius $r_{min}$; negative selection occurs with a given large probability, typically $p_N=0.99$.

\emph{Positive selection} kills pre-Th cells that are \emph{farther from each self peptide} than a maximum radius $r_{max}$; positive selection occurs with a given, relatively smaller probability, typically $p_P=0.9$. This way, some of the randomly generated Th cells that cannot bound self-peptides may still survive and they can become infection or mutation specific Th cells later.

The \emph{degree of maturity} of a na\"{\i}ve Th cell is 0. If a TCR is in the \emph{characteristic ring} around the reflected image of some self-peptide (see Fig.~\ref{fig:math_model}A), that is, $r_{min} < d(z_P, \overline{z_T}) < r_{max}$, then it is called a \emph{regulatory Th cell} and its degree of maturity is set to 2. Here $\overline{z_T} := (x_T, -y_T)$ is the center of the recognition region of the Th cell and $z_P := (x_P, y_P)$ represents the shape of a self-peptide. In our model a regulatory Th cell has double role. On one hand, it takes part in the controlling role of the regulatory T cell repertoire, but it can also act as a Th cell.

Other Th cells that have survived the negative and positive selections, but are outside of the characteristic ring of each self-peptide, are called \emph{potential infection or mutation specific Th cells}, and their degree of maturity is set to 1.

When simulating a CRS model, we set $r_{min}:=r_{max}$. Thus positive selection and this type of regulatory Th cells are not simulated then. Care is taken that the average number of Th cells be the same as in the case of ERS model, so the number of potential infection specific Th cells are larger in the CRS model than in the ERS model.

\paragraph*{Th cell actions}
For each Th cell, there is a sequence of actions, with exponential random waiting times between two actions. At each action the Th cell is to randomly choose one of the potential target MHCII+peptide complexes in its recognition region. The closer an MHCII+peptide complex to the center of the recognition region, the bigger its chance of being selected.

\paragraph*{Th cell activation control process} \label{sssec:Tstress}
It is a sequence of frequently occurring random events whose purpose is to check and possibly change the \emph{state of activation} of a Th cell. A Th cell can be in a state of \emph{activated} or \emph{non-activated}. This process checks if this Th cell has received danger signal in a critical period of time before this check. If the result of this check is ``yes'', then the Th cell is set to ``activated'' (stress=1); otherwise it is set to ``non-activated'' (stress=0).

An ``activated'' Th cell starts an interleukin secreting process. This process is a signal of its activated state for ``activated'' B cells in its environment. We use the symbolic names ``interleukins'' in this paper, without specifying the exact type of these interleukins.

An ``activated'' Th cell begins \emph{cell division of intermediate kind}. Division of the intermediate kind is different from the weak or strong kind. In the CRS model we use only division of the strong kind.

\paragraph*{Self-nonself discrimination} \label{sssec:self-nonself}
It is important that in the ERS model, self-nonself discrimination is solved by the \emph{complete repertoire of Threg cells}. When a \emph{regulatory} Th cell (degree of maturity is 2) bounds with intermediate affinity a B cell's MHCII+peptide complex which has state ``non-activated'', then with high confidence it means that the peptide is a \emph{self-peptide}. This contact initiates a \emph{division of weak kind} for both this regulatory Th cell and the attached B cell. This weak division helps to stabilize this interaction among three partners: self-cells, B cells that can react to self, and regulatory Th cells that can attach to this self-peptide with intermediate affinity. It is important that B cells that can contact Threg cells with all their MHCII-peptide complexes cannot start an intermediate or strong division process. It gives the important \emph{inhibitory effect} of Threg cells. This way, B cells that react to self are in a state of ``non-activated'' permanently with large probability.

When a Th cell that has already went through the thymus, obtains danger signal then it may begin a non-specific \emph{division of intermediate kind} and may start to secrete interleukins to start \emph{division of intermediate kind of activated B cells}.

If a Th cell has already went through the thymus, but it is not a regulatory Th cell (thus its degree of maturity is 1), the target is an activated B cell, and the distance of attachment satisfies $d(z_P, \overline{z_T}) < \frac{r_{min}}{2}$, then with high confidence it means that the peptide is \emph{foreign or mutated self}. Here $\overline{z_T} = (x_T, -y_T)$ is the center of the recognition region of the Th cell, $z_P := (x_P, y_P)$ is the point representing the peptide, and $r_{min}$ is the inner radius of the characteristic ring around the reflected image of self-peptides. Remember that because of the negative selection, such short distance between a self-peptide and the center of recognition region is extremely unlikely. Then both this B cell and Th cell are very likely useful tools to fight against an infection. As a result, this interaction may initiate a \emph{division of strong kind} both in the affected B and Th cells, plus stimulates the secretion of danger signal (in the B cell) and interleukins (in the Th cell). Strong division of a B cell implies its hypermutation with given probability as well. This is a \emph{direct help} of the Th cell for the affected B cell.

\paragraph*{Th cell divisions} \label{sssec:Thdiv}
The probability of division of a Th cell may depend on several factors. It may get bigger when the distance $d$ between the MHCII+peptide complex and the TCR is smaller (i.e., the complementarity is better). It gets smaller when the number $n_0$ of all TCR's is large (i.e., the concentration of Th cells is already large). It gets smaller when the number $n_1$ of TCR's in a neighborhood of the Th cell is large (i.e., the local concentration of Th cells is already large). The formula for the probability of division is given by a somewhat different formula for the strong division; namely, weak and intermediate divisions do not depend on the complementarity distance $d$. The reason is that weak reaction by definition have relatively uniform distance between Threg cells and a self antigen, see the rings of the ERS model in Fig.~\ref{fig:math_model}B. Also, intermediate reactions are by definition non-specific, with almost arbitrary distance $d$. When simulating CRS models, we use only strong division.

The probability of a \emph{division of weak kind} of a Th cell is given by
\begin{equation}\label{eq:Thdivw}
p_{T,w} = k_w \, g_{\theta_{n0}, \eta_{n0}}(n_0) \, g_{\theta_{n1}, \eta_{n1}}(n_1) .
\end{equation}
The purpose of division of weak kind is to establish a stable contact between self antigens, B cells reacting to self with a weak affinity, and Threg cells reacting to self peptides with an intermediate, standard affinity.

The probability of a \emph{division of intermediate kind} of a Th cell is given by
\begin{equation}\label{eq:Thdivm}
p_{T,m} = k_m \, g_{\theta_{n0}, \eta_{n0}}(n_0) \, g_{\theta_{n1}, \eta_{n1}}(n_1) .
\end{equation}
The purpose of division of intermediate kind is to create a fast, non-specific immune reaction to a new, typically quickly growing number of nonself antigens. The growing amount of Th cell help (interleukins) can help the division of intermediate kind of B cells that are able to bind the new nonself antigens in the humoral phase.

The probability of a \emph{division of strong kind} of a Th cell is given by
\begin{equation}\label{eq:Thdivs}
p_{T,s} = k_s \, g_{\theta_d, \eta_d}(d) \, g_{\theta_{n0}, \eta_{n0}}(n_0) \, g_{\theta_{n1}, \eta_{n1}}(n_1) .
\end{equation}
The purpose of division of strong kind of Th cells is to initiate a strong immune reaction when infection or mutation specific Th cells appear and can bind infection or mutation specific B cells. Important requirements to such a division that the binding distance satisfy $d < \frac{r_{min}}{2}$ and the attached MHCII be ``activated''. These requirements can guarantee with large probability that this strong reaction is not arising against self. Then this Th cell becomes ``strongly activated'' (stress=2). This condition is independent of danger signal.

For simplicity, the constants $k_w, k_m, k_s$ above are typically set to $1$.

\paragraph*{Regulatory Th cells} \label{sssec:Threg}
As we saw above, the regulatory Th cell repertoire plays a most important controlling role in self--nonself discrimination in our ERS model. When simulating CRS models we do not use Thregs explicitly, since their conventional role is to prevent autoimmunity, and when we compare the ERS and CRS models, autoimmunity is avoided. Their role in the ERS model is similar to the one they have in the computational model~\cite{LLC03}. Starting in the fetus, and throughout the entire life span, they give a faithful mirror-image of the self-peptide repertoire.
\begin{itemize}
  \item They regularly visit B cells having only self-peptides on their MHCII and inhibit their strong division, but support their weak division.
  \item They are players in normal Th cell roles, like helping non-specific intermediate type and specific strong type division of B cells. They can also secrete interleukins.
\end{itemize}

\subsection{B cells} \label{ssec:B}

Na\"{\i}ve B cells are born in the bone marrow. The birth of a B cell initiates its own (natural) death event, B cell action process, and B cell activation control process, each with separate rate. Each B cell carries a number (say, 3) of MHCII molecules.

\paragraph*{B cell recognition region}
Each B cell has a recognition region in the antigen lattice. If a BCR is described by the point $(x_B, y_B)$, then the corresponding recognition region is a square with center $(x_B, -y_B)$ and radius $r_B$. The radius of the BCR of a na\"{\i}ve B cell is a given constant, while B cells that are born in the periphery after hypermutation may have smaller radii. The BCR $z'_B=(x'_B, y'_B)$ of a hypermutated B cell offspring is determined at random, uniformly on a square around the mother BCR. \emph{Thus there is only a chance that its affinity to a given antigen} $z_A=(x_A, y_A)$ \emph{is higher}, that is, the distance $d(z_A, \overline{z'_B})$ is smaller than that of its mother cell. The radius $r'_B$ of a hypermutated offspring will be smaller than that of its mother cell depending on the above distance: $r'_B = c \, r_B + r_0$. Typical values are $c=0.9$ and $r_0=5$. This effect may increase the affinity of some ``lucky'' offspring to the given antigen.

In sum, the recognition region describes the potential shapes of antigens with which a BCR can bind: the smaller the distance between an antigen $z_A=(x_A, y_A)$ and the center $\overline{z_B}=(x_B, -y_B)$ of the recognition region, the better the fit between the antigen and the BCR.

\paragraph*{B cell action process} \label{sssec:Bact}
For each B cell, there is a sequence of actions, with independent exponential waiting times between two actions. At each action the B cell is to randomly choose one of the potential target antigens in its recognition region. A target can be another B cell, an antibody, a non-immune self cell, or a foreign antigen.
The closer an antigen $z_A=(x_A,y_A)$ to the center $\overline{z_B}=(x_B, -y_B)$ of the recognition region, the bigger its chance of being selected as the next target. The chosen target can be killed only if the above distance is smaller than the recognition radius $r_B$ of the B cell, that is, $d(z_A, \overline{z_B}) < r_B$. The smaller this distance, the larger the probability that the antigen will really be destroyed. Since smaller distance represents stronger affinity in the model, it means longer attachment between an antigen and the BCR. So this condition is equivalent to the fact that a target can be killed if it is bound to the BCR for a long enough time.

When the chosen target is destroyed, its peptide is placed on one of the MHCII's of the B cell. The MHCII selected is primarily an empty one; when all of the MHCII's are already loaded, then one of them is chosen at random to replace the old peptide by the new one.

\paragraph*{B cell negative selection filter in the bone marrow}
To each na\"{\i}ve (immature) B cell there is assigned a random event that places it into a negative selection filter in the bone marrow. Negative selection kills B cells that are \emph{closer to one of the self-antigens} than a minimum radius $r_{minb}$; negative selection occurs with a given large probability, typically $p_{Nb}=0.99$.

The \emph{degree of maturity} of a na\"{\i}ve B cell is 0. A B cell that has survived the negative  selection is called a \emph{mature B cell}, and their degree of maturity is set to 1. Only B cells with degree of maturity $\ge 1$ can function as normal B cells.

\paragraph*{B cell activation control process} \label{sssec:Bstress}
In the ERS model, it is a sequence of frequently occurring events whose purpose is to check and possibly change the \emph{state of activation} of a B cell. It is not used in the CRS models. The main parameter is the \emph{critical time} $t_{crit}$. Each of the MHCII carried by a B cell can be in a state of \emph{``activated''} or \emph{``non-activated''}. An empty MHCII is not ``activated'' by definition.
\begin{itemize}
  \item A given non-empty MHCII is set to ``non-activated'' when the time elapsed since the last event effecting this MHCII is less than $t_{crit}$. Such an event can be a regulatory Th cell attaching to this MHCII, or placing a new peptide on this MHCII.
  \item A given MHCII is set to ``activated'' when the time elapsed since the last event effecting this MHCII is greater than or equal to $t_{crit}$.
\end{itemize}

Similarly, a B cell can also be in a state of \emph{``activated''} or \emph{``non-activated''}.
\begin{itemize}
  \item When its each MHCII is in the state of ``non-activated'', the B cell itself is set to state of ``non-activated''.
  \item When at least one of its MHCII is ``activated'', then the B cell is set to ``activated''.
\end{itemize}

An ``activated'' B cell starts an \emph{danger signal} sending process. This process is a signal of its activated state for Th cells in its environment.

An ``activated'' B cell may start a \emph{cell division of intermediate kind} if it obtains help from non-specific Th cells. Help may come as interleukins produced by Th cells, that has arrived in a critical period of time before this check. (This kind of cell division cannot occur with plasma cells or memory cells.) Division of the intermediate kind is different from the weak or strong kind. Here the activation (stress) level is 1.

In the case of \emph{cell division of the strong kind}, which occurs by the help of infection or mutation specific Th cells, the activation (stress) level is 2.

\paragraph*{B cell division and maturity} \label{sssec:Bmat}
Each B cell has \emph{a degree of maturity}. A na\"{\i}ve, immature B cell has degree 0, while B cells that have survived a negative selection filter in the bone marrow are mature B cells, having degree of maturity 1 first. Mature B cells may encounter antigens at the periphery. A B cell division can be the result of an encounter with an antigen which is escorted by a direct or indirect (via interleukin) help from a Th cell. At each division of a B cell, one of the two offspring inherits all characteristics of the mother cell (let us call it the \emph{first offspring} for explicitness), while the other offspring (let us call it the \emph{second offspring}) may undergo hypermutation with given probability. The first offspring inherits the mother's MHCII-peptide complexes, while the second offspring starts with MHCII molecules with a default (non-specific) peptide. The second offspring after the first division has a degree of maturity 2. The result of a hypermutation is a B cell with randomly shaped BCR. The possible shapes are uniformly distributed on a square of the antigen lattice, with given radius around the mother BCR.

A second division may lead to two different outcomes with given probabilities: the second offspring can be either a \emph{memory cell} (degree=3) or a \emph{plasma cell} (degree=4). A memory cell has the same characteristics as a normal B cell except that its average lifespan is significantly longer (e.g 10 days instead of the standard 3 days). A plasma cell constantly -- at random time instants -- produces antibodies of the type of its own BCR.

Possibility of division of a B cell arises after contacting an antigen or obtaining Th help in the form of interleukins. The probability of division of a B cell depends on several factors. It gets bigger when the distance $d$ between the antigen and the BCR is smaller (i.e., the complementarity is better), or when the radius $r$ of the recognition region of the BCR is smaller (i.e., the affinity of the B cell is bigger). It gets smaller when the number $n_0$ of other BCRs in a rectangle around the BCR is small (i.e., the concentration of B cells is already large). Finally, one or two factors can depend on the concentration difference $c$ between the number of targets in the recognition region of the B cell and the number of targets in the reflected image of the recognition region. If the concentration difference is too small, the B cell may get insensitive. If the concentration difference is too large, the B cell may get anergic.

In the ERS model, the specific formulas for the probability of division in the respective cases of weak, intermediate, and strong B cell divisions are as follows. In the CRS models only the strong division is used. The probability of a \emph{division of weak kind} of a B cell is given by
\begin{equation}\label{eq:Bdivw}
p_{B,w} = k_w \, g_{r_{mb}+\theta_d, \eta_d}(d) (1 - g_{r_{mb}-\theta_d, \eta_d}(d)) \,g_{\theta_r, \eta_r}(r) \, g_{\theta_n, \eta_n}(n_0) \, (1 - g_{n_m, \eta_{c}}(c)) .
\end{equation}
The purpose of division of weak kind is to establish a stable contact between self antigens, B cells reacting to self with a weak affinity, and Threg cells reacting to self peptides with an intermediate, standard affinity. The first, constant factor $k_w$ is typically 1. The purpose of the second and third factors depending on $d$ is to help those B cells that are at a standard distance from their targets, in the present case, self antigens. The last factor, depending on $c$ intends to guarantee that a large number of antigens, typical for self antigens, be in the recognition region of the weakly dividing B cells. The first parameter $n_m$ here is the actual number of bone marrow cells, which is a common measure of the size of non-immune self cell populations.

The probability of a \emph{division of intermediate kind} of a B cell is given by
\begin{equation}\label{eq:Bdivm}
p_{B,m} = k_m \, g_{\theta_d, \eta_d}(d) \,g_{\theta_r, \eta_r}(r) \, g_{\theta_n, \eta_n}(n_0) \, g_{\theta_{c2}, \eta_c}(c) \, (1 - g_{\theta_{c1}, \eta_{c}}(c)) .
\end{equation}
The purpose of division of intermediate kind is to create a fast, non-specific immune reaction to a new, typically quickly growing number of nonself antigens. The growing amount of B cells that are able to bind the new nonself antigens in the humoral phase even when there exist no infection or mutation specific B or Th cells can give an early start to an effective immune reaction. Activated B cells can release danger signal to initiate a non-specific Th help as well. The value of the constant multiplier $k_m$ is typically $100$ to create a fast answer to a new, quickly dividing infection.

The probability of a \emph{division of strong kind} of a B cell is given by
\begin{equation}\label{eq:Bdivs}
p_{B,s} = k_s \, g_{\theta_d, \eta_d}(d) \,g_{\theta_r, \eta_r}(r) \, g_{\theta_n, \eta_n}(n_0) \, g_{\theta_{c2}, \eta_c}(c) \, (1 - g_{\theta_{c1}, \eta_{c}}(c)) .
\end{equation}
The purpose of division of strong kind of B cells is to initiate a strong immune reaction when infection or mutation specific Th cells appear and can bind infection or mutation specific B cells. Important requirements to such a division that an ``activated'' Th cell binds an ``activated'' MHCII of this B cell and the binding distance between the reflected image of the TCR and the peptide is smaller than $\frac{r_{min}}{2}$. These requirements can guarantee with large probability that this strong reaction is not arising against self. The value of the constant $k_s$ is typically $200$ to create a strong reaction when -- tipically -- the number of B cells specific to a new infection is very low.

\paragraph*{B cell affinity maturation and network memory}
Like in natural selection, there exists neither intelligent control which would direct genetic mutations toward better fit, nor memory that would save cells from genetically searching a proved wrong ``direction''. The major effect which has physiological consequences on a B cell is the strength of antigen binding. This is like finding the source of heat in a dark room, using a single thermometer, with no direct sensing of direction and with no memory. The technique the present model applies is a microscopic analog of evolution: hypermutation and selection, with survival of the fittest. Namely, the program uses a stochastic search for best fit (or a stochastic learning process):
\begin{itemize}
\item An offspring may be randomly hypermutated, so a random variation is created in the affinity to the given antigen.

\item The stronger a B cell can bind a given antigen, the more offspring it can produce.

\item When the concentration of the given antigen is decreasing, a competition arises among B cells for the antigen, and those having higher affinity would win in this selection process.
\end{itemize}

An affinity maturation model has to handle the danger of autoimmunity. Even if na\"{\i}ve Th cells which can strongly bind self peptides are deleted as a result of negative selection in the thymus, and also na\"{\i}ve B cells which can strongly bind self antigens are deleted as a result of negative selection in the bone marrow, still there is the danger that autoimmune B cell clones may be produced as a result of hypermutation. In the presented model there is a double defense against this danger.
\begin{itemize}
\item The absence of T cell help in the case of B cells that react strongly to non-immune self antigens inhibits their division. This is an essential difference between self and nonself in the model.
\item Since nonself antigens which can start somatic hypermutation typically appear after birth, when the number of self cells is already very large, one can argue that at that time randomly produced self-reactive B cell clones are confronted with an overwhelming quantity of self antigens. As a result, these B cell clones would become anergic~\cite{GCJ89}. In the model this is simulated in the B cell division process: divisions of a B cell, see (\ref{eq:Bdivm}) and (\ref{eq:Bdivs}), become less frequent when the number of objects in its recognition region becomes overwhelmingly large. The reproduction process of B cells is fastest when the concentration of the complementary antigens is neither too small, nor too large. This is common for both self or nonself antigens in the model, so when nonself overgrows an upper threshold, the model immune system remains practically defenseless against it as well.
\end{itemize}

As a result of the double defense described above, there will be ``holes'' in the adaptive immune system, both in the T cell and B cell populations, around the mirror image of non-immune self cells~\cite{GCA88,GAB90}. The negative selection in the model is especially important during early ontogenesis when the smaller population of host cells is vulnerable to self-reactive immune cells. As the individual reaches adult size, the large number of host cells plus the absence of T cell help can alone inhibit reproduction and affinity maturation of immune cells. Then negative selection in the model (like in reality in the thymus) becomes less essential.

It is reasonable to expect that after a somatic hypermutation –- affinity maturation process the resulted specific B cell clones may survive for a certain period of time as a local memory. In the model, expansion of certain B cell clones (e.g. as a result of an infection by a foreign antigen), under favorable conditions, stimulates the reproduction of secondary B cells which are complementary to the expanded primary B cell clones and whose receptors are, therefore, similar to the infecting antigens. (Of course, similarity here means a mimicry of a binding partner and not similarity at the molecular level.) Thus a mirroring process (``ping-pong'') and a local network memory may develop and last for a longer time, even in the absence of the stimulating antigen. While this memory lasts, repeated infection of the same pathogen is eliminated more efficiently. This network model of immune memory essentially conforms to Jerne's immune network concept~\cite{Jer74}. Beside other factors, like longer living memory cells or antigen preserving follicular dendritic cells, this could be a possible explanation of immune memory.

\subsection{Antibodies} \label{ssec:Anti}

A plasma cell is a special kind of B cells, a result of a B cell maturity process. A plasma cell has neither a B cell action process, nor a B cell activation control process. On the other hand, it has an antibody birth and an antibody death process. An antibody has the same shape in the antigen lattice as the BCR of its mother plasma cell.

Antibodies have similar action processes as B cells, but, naturally, when tagging a target, peptide of the target does not appear on an MHCII. The complement sub-system of the immune system is currently not represented in the model, so it is supposed that when an antigen is tagged by an antibody, it leads to the destruction of the targeted antigen with a certain probability.

\subsection{Foreign antigens} \label{ssec:Foreign}

After birth, different pathogens may enter the body, perhaps several times (e.g. repeated infections with the same pathogen). A foreign antigen is represented by its position $(x_F, y_F)$ in the antigen lattice and its peptide $(x_{FP},y_{FP})$ in the peptide lattice. A foreign cell comes with an initial population size and a birth process with a given initial rate (that is, with a given initial average waiting time $\tau_{f0}$ between divisions). If the size of the population of a specific pathogen at a certain time $t$ is $f=f(t)$, then the conditional expected waiting time between two divisions in this population is
\begin{equation}\label{eq:foreignbirth}
\tau_f = \frac{\tau_{f0}}{f \, g_{\theta,\eta}(f)} = \frac{\tau_{f0}}{f} \left(1 + \left(\frac{f}{\theta}\right)^{\eta}\right).
\end{equation}
For the sake of simplicity, the natural death process of foreign antigens is not represented in the model. So (\ref{eq:foreignbirth}) should be more accurately called the effective growth process of foreign antigens.

\section{Simulation results of MistImm} \label{sec:Sim}

In this chapter we assign specific parameter values for the MiStImm computational model and describe the results obtained by computer simulations. 
MiStImm can be initialized by approximately one hundred input parameters (Table \ref{tab:input parameters} and \ref{tab:input parameters_c}.)
The parameters can be used to set various immune system models, including the above mentioned ERS and CRS models. 
Once an immune system model is fixed, further individual settings are available (for example, foreign cell injections with different numbers or types). 
Different initial random numbers can be set to run different random realizations with the same parameter settings. 

In the following, we show that a typical simulation by the ERS model fits the basic requirements that are expected from an immune system model. 
Then we compare the simulation results of the ERS and the CRS models. 

\begin{table}[ht!]
\footnotesize
\centering
\caption{The most important parameters of the MiStImm computational model. The unit of time is one-tenth of a day: 2 hours and 24 minutes.}
\label{tab:input parameters}
\begin{tabular}{|l|l|c|}
\hline
\textbf{Parameter} & \textbf{Description} & \textbf{Typical value} \\ \hline

comptype & computational model & 0: ERS, 1: CRS \\

medrepr &  intermediate interaction & 0: off, 1: on \\

weakrepr & weak interaction & 0: off, 1: on \\

nrmax & the simulation stops at this number of pathogens & 5000 \\

nm & initial number of bone marrow cells & 5 \\

timmst & starting time of the immune system & 100 \\

tmax & the last time instant of the simulation & 5000 \\

xmax & size of the antigen lattice & 1000 \\

r0 & radius of na\"{i}ve B cells & 140 \\

r0s & radius of spreading area of offspring B cells & 60 \\

tlifeb & mean life length of a B cell & 30 \\

tlifmem & mean life length of a memory B cell & 150 \\

pmem & the probability of B cell changing into memory cell & 0.3 \\

thrad & radius of Th cells & 80 \\

pxmax & the size of the peptide lattice & 1000 \\

rminth & threshold radius of negative selection & 30 \\

rmaxth & threshold radius of positive selection & 50 \\

rminb & threshold radius of negative selection of B cells & 140 \\

pmut & the probability of B cell hypermutation at reproduction & 0.4 \\

taum & mean time between two divisions of a bone marrow cell & 400 \\

taubm & mean time between two births of B cells in b. marrow & 30 \\

tauselb & mean time for a B cell to enter negative selection & 0.05 \\

taub & mean time between two actions of a B cell & 5 \\

taubstress & mean time between two B cell activation checking & 0.5 \\

taubil & mean time between the births of IL type 2 & 0.2 \\

taudil & mean time between the deaths of IL type 2 & 30 \\

tauab & mean time between two actions of an antibody & 0.5 \\

taubab & mean time between two births of antibodies & 1 \\

taudab & mean time between two deaths of antibodies & 80 \\

tauthm & mean time between two births of Th cells in b. marrow & 5 \\

tauthymus & mean time for a Th cell to enter the thymus & 0.05 \\

tauth & mean time between two actions of a Th cell & 2 \\

tlifeth & mean life length of a Th cell & 30 \\

sreprcrit & threshold radius of the strong reproduction & 40 \\

dring & radius within which the cc. of Th cells are restricted & 10 \\

tcritilb & crit. time between arrivals of two IL type 2 at a B cell & 1 \\

tcritth & crit. time between two Th cells arrival at a given MHCII & 2 \\

tauprodil1 & mean time between two births of IL type 1 & 0.2 \\

taudil1 & mean time between two deaths of IL type 1 & 30 \\ 

tauil1 & attack rate of an IL type 1 & 1 \\

tauil & attack rate of an IL type 2 & 1 \\

tthcrit & crit. time between arrivals of two IL type 1 at a Th cell & 1 \\

tauthstress & rate of the Th activation control process & 0.5 \\

negselp & threshold probability of Th negative selection & 0.99 \\

posselp & threshold probability of Th positive selection & 0.9 \\

bselp & threshold probability B cell negative selection & 0.99 \\

th* & different theta parameters, see (\ref{eq:logisticf}) & 1--1000 \\

eta* & different eta parameters, see (\ref{eq:logisticf}) & 1--4 \\ 
\hline

\end{tabular}

\end{table}

\begin{table}[ht!]
\footnotesize
\centering
\caption{The important parameters of the MiStImm computational model, continued. The unit of time is one-tenth of a day: 2 hours and 24 minutes.}
\label{tab:input parameters_c}
\begin{tabular}{|l|l|c|}
\hline
\textbf{Parameter} & \textbf{Description} & \textbf{Typical value} \\ \hline

kth0, 1, 2  & multipliers of Th reproduction, see (\ref{eq:Thdivw}), (\ref{eq:Thdivm}), (\ref{eq:Thdivs}) & 1 \\

kb0 & multiplier of B weak reproduction, see (\ref{eq:Bdivw}) & 1 \\

kb1 & multiplier of B intermediate reproduction, see (\ref{eq:Bdivm}) & 100 \\

kb2 & multiplier of B strong reproduction, see (\ref{eq:Bdivs}) & 200 \\

nwtypes & number of types of self cells & 1--4 \\

nw & initial number of a specific self cell & 150 \\

xw & x-coordinate of a specific self cell & 0--1000 \\

yw & y-coordinate of a specific self cell & -500--500 \\

t0w & starting time of a self cell type & 0 \\

tauw & mean time between two divisions of a spec. self cell & 40 \\

nrtypes & number of types of pathogens & 1--4 \\

nr & initial number of a specific pathogen & 200--700 \\

xr & x-coordinate of a specific pathogen & 0--1000 \\

yr & y-coordinate of a specific pathogen & -500--500 \\

t0r & starting time of a specific pathogen & 3000--4000 \\

taur & mean time between two divisions of a spec. pathogen & 30--80 \\ \hline

\end{tabular}

\end{table}

\subsection{Development and homeostasis of the immune system} \label{ssec:Develop}

\begin{figure}[ht!]
     \centering
     \includegraphics[width=\linewidth]{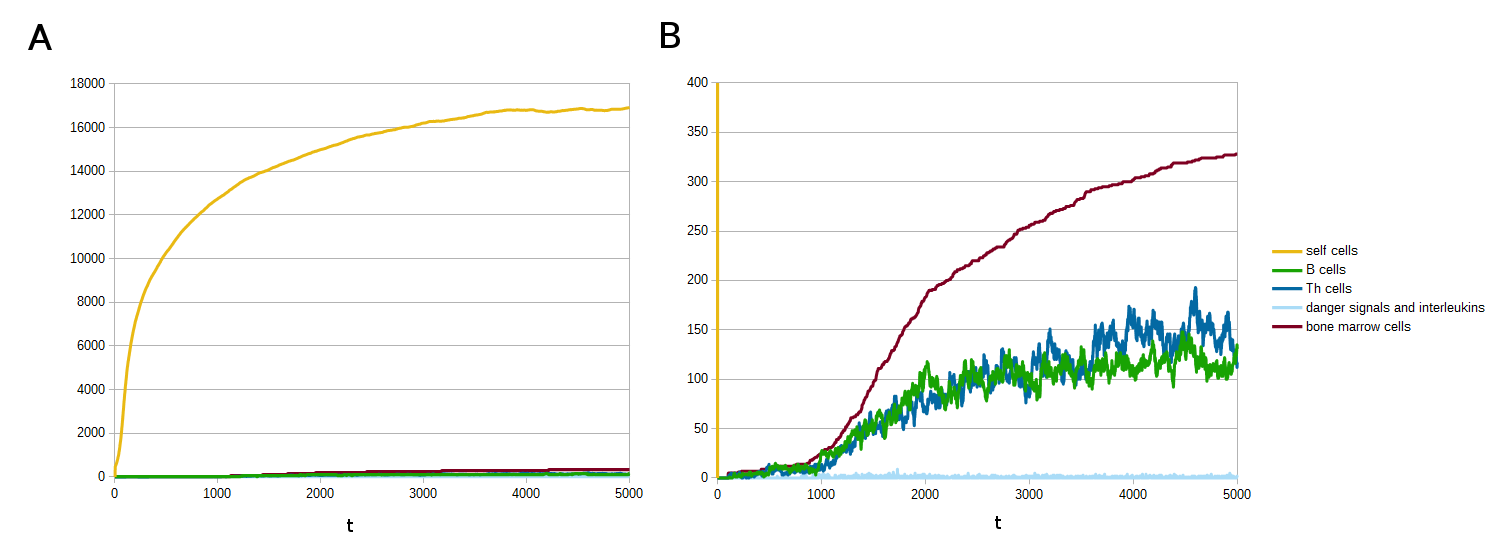}
     \caption{Development and homeostasis of the immune system. The same single simulation in bird's eye view ({\bf A}) and a closer view ({\bf B}), respectively. Horizontal axis: time ({\it t}) from conception measured in one-tenths of a day. Vertical axis: number of cells/molecules. In the case of self cells the sum of sizes of the self cell populations is displayed.}
     \label{fig:dev_and_homeo}
\end{figure}

A simulation starts from a few days after conception and goes until the $5000$th time instant; the unit of time being a tenth of a day ($2$ hours and $24$ minutes). Initially, only three types of non-immune self cell populations appear in the model, each with a number of $150$ cells, and no other components. 
Each of these populations is accompanied by a cell division process that implies continuous growing of the number of self cells, with decreasing rate in time (Fig.~\ref{fig:dev_and_homeo}A). 
B and T cells, which generated by the bone marrow cells, first appear at the $100$th time instant (Fig.~\ref{fig:dev_and_homeo}B). The number of these cells also grow continuously with a decreasing rate.

\begin{figure}[ht!]
     \centering
     \includegraphics[width=\linewidth]{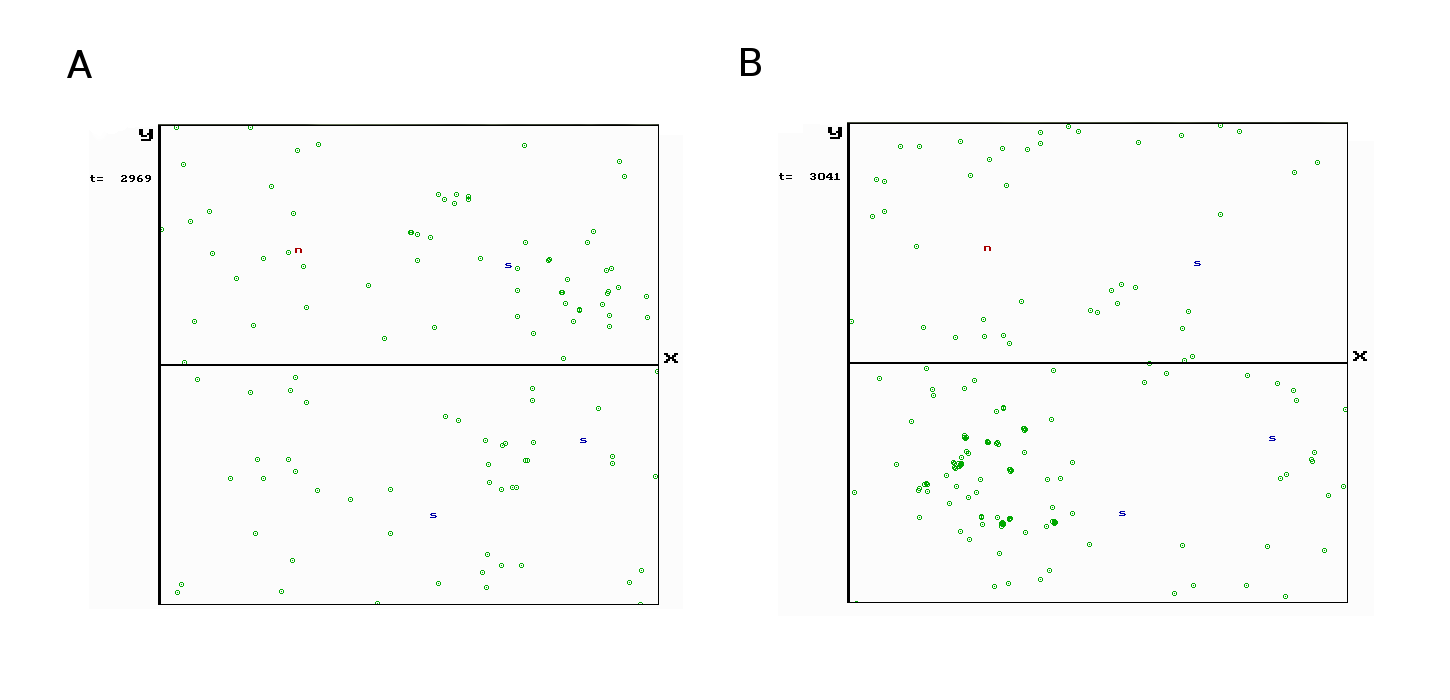}
     \caption{Peptide and antigen lattices in the ERS model. ({\bf A}) A snapshot of the peptide lattice, where the actual TCRs are displayed. With random `rings' around the reflected images of non-immune self antigens (`s') about one month after birth. A movie capturing a typical simulation of the peptide space is available at the address https://goo.gl/QcdG48. ({\bf B}) A snapshot of the antigen lattice, where the actually existing BCRs are displayed. B cell response to a pathogen: large density of pathogen specific B cells at the reflected image of nonself (`n') about one week after the infection. As a result of negative selection, there are empty domains around the reflected images of non-immune self antigens (`s'). A movie capturing a typical simulation of the antigen lattice (shape space) is also available at the address https://goo.gl/3oK1bM.}
     \label{fig:lattices}
\end{figure}

The peptide and the antigen lattices both have size $\{0,1000\}\times\{-500,500\}$.  
Coordinates of antigens of three different self cells (denoted by letters `s') are $(550,300)$, $(700,-200)$ and  $(850,150)$, both in the case of the peptide and the antigen lattice (Fig.~\ref{fig:lattices}AB). 
TCR rings around the mirror images of self peptides -- that are characteristic features of the ERS model -- begin to develop about the $1500$th time instant and become more or less stabilized by the $2800$th time instant (Fig.~\ref{fig:lattices}A). 
These rings fluctuate for two reasons: {\it (i)} occasionally global Th cell populations overgrow the set upper limits and this reduces the probability of Th cell division; {\it (ii)} sometimes the number of presented self peptides in the MHCII-peptide complexes of B cells reaches an extremely low level.
An infection brings significant changes. A rising population of B and Th cells appear at the mirror image of the infecting agent denoted by a letter `n' (Fig.~\ref{fig:lattices}B). 

\subsection{Immune response} \label{ssec:Normal}

An immune system do not attack self cells strongly, just to a very limited extent. Some B cells must continuously present self peptides to ensure that Threg cell characteristic rings around self peptides are constantly maintained. Because of negative selection, this type of immune response is weak and typically settles down quickly before it becomes pathological.
An immune response should have the ability to destroy the majority of pathogens -- some of them suddenly, others perhaps slowly, while in some cases it may fail. In ERS model, death of an individual occurs when the pathogen population grows up irreversibly, technically, as its size reaches $4000$ cells. Diversity of pathogens are represented by different locations of their receptors, different speeds of growth, and different initial numbers.

\begin{figure}[ht!]
     \centering
     \includegraphics[width=\linewidth]{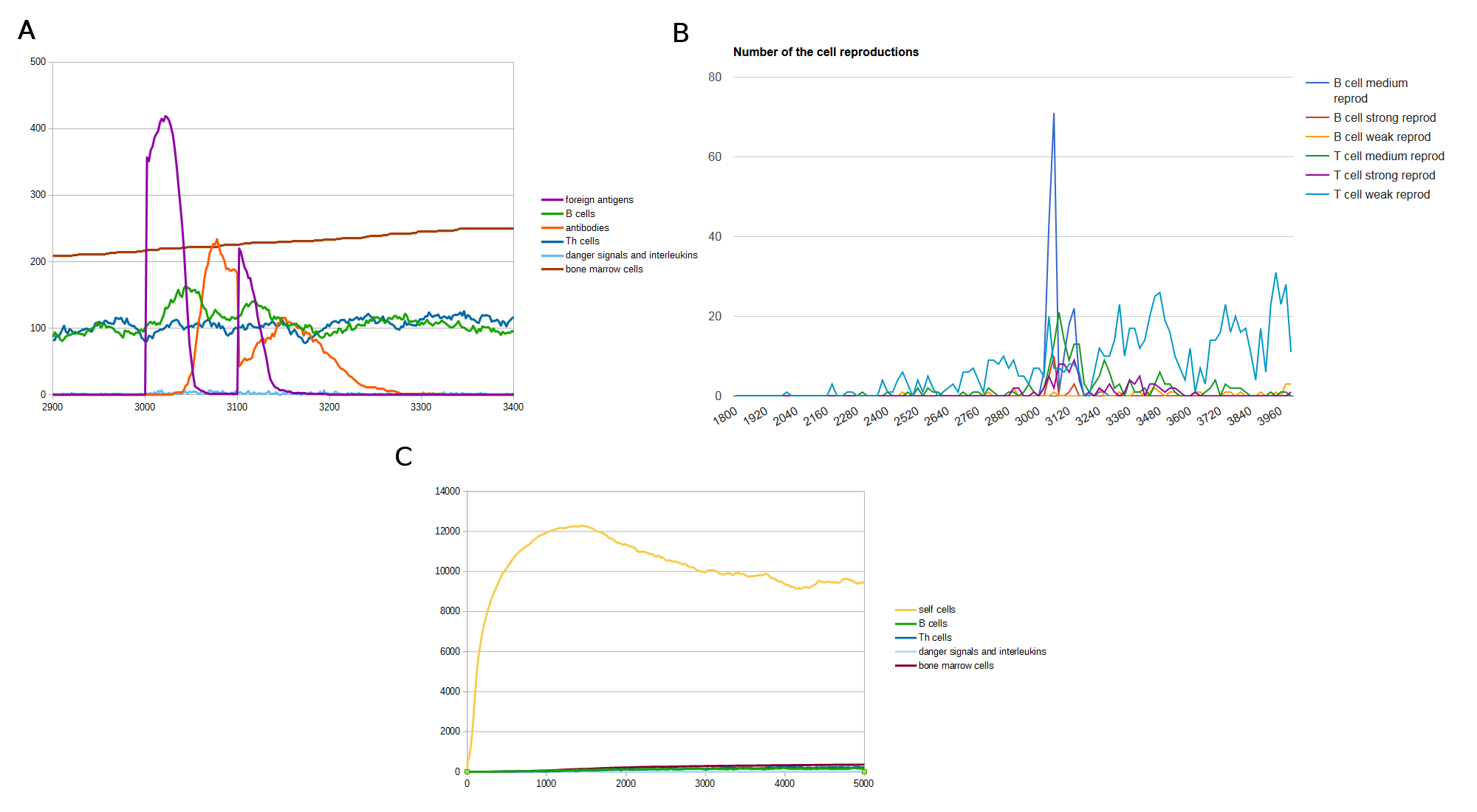}
     \caption{Immune response in the ERS model. ({\bf A}) Normal immune response against a repeated infection. First infection was injected at the time instant $3000$ and the second infection was injected at the time instant $3100$. Both infections are the same type (number of cells $= 350$, mean waiting time between two divisions $= 60$). The second infection was eliminated much faster because of B cell memory. Horizontal axis: time from conception measured in one-tenths of a day. Vertical axis: number of cells/molecules. ({\bf B}) Division of weak/intermediate/strong kind of T and B cells in the same simulation showed in ({\bf A}). ({\bf C}) Autoimmunity caused by the lack of negative selection of B cells.  Horizontal axis: time from conception measured in one-tenths of a day. Vertical axis: number of cells/molecules.}
     \label{fig:screenshots}
\end{figure}

A normal immune response develops immune memory. Thanks to memory cells, a second immune response against the same nonself antigen is more effective than at a primary infection (Figure~\ref{fig:screenshots}AB). To test this phenomenon, we have performed 500 simulations, adding the same type of pathogen (number of cells $= 350$, mean waiting time between two divisions $= 60$) at the 3000th and at the 3150th time instants. ERS model have won against both infections in 451 cases and the mean time lengths needed for elimination were 62.02 (std  13.26) at the first infection and 20.51 (std 14.94) at the second infection. 
We have said that an elimination happened when the number of pathogens have decreased under 50. With two-tailed t-test, the $p$-value for equality of mean elimination times for the first and the second infection was $5.2e-227$. 

Experiments with our computational model showed that immune system  cannot fight effectively against more than a couple simultaneous infections. Similar is true in the case of the development of immune memory. This observation fits experimental results~\cite{XKXL15}.

Lack of negative selection of B cells results autoimmunity (Fig.~\ref{fig:screenshots}C). Without negative selection, some of the B cells can constantly destroy self cells. 

\subsection{ERS vs. CRS model} \label{ssec:ERS_CRS}

One can switch the ERS (\emph{Enhanced Role of Self}) model to a CRS (\emph{Conventional Role of Self}) model by modifying four parameters. 
Turning off the {\it division of weak type} and the {\it division of intermediate type} are required in the CRS model (medrepr $= 1 \rightarrow 0$ and weakrepr $= 1 \rightarrow 0$). 
Turning off positive selection of T cells is also required in the CRS model (comptype $= 0 \rightarrow 1$). 
The latter adjustment causes large growth of the T cell population, so simultaneously we need to decrease the expectation of the waiting time between two births of T helper cells in the bone marrow (tauthm $= 5 \rightarrow 30$).

We compared the efficiencies of the immune reactions in the two models. 
Our results showed that in the ERS model the adaptive immune reaction was able to destroy infections with critically large initial numbers or with critically fast division times more often than in a CRS model (Table~\ref{tab:sim_res_1} and \ref{tab:sim_res_2}). Fisher's exact test was used for the statistical evaluation (Table~\ref{tab:Fisher}). All the corresponding simulation results can be seen at the address \url{http://info.ilab.sztaki.hu/~kerepesi/MiStImm/}.

\begin{table}[ht!]
    \centering
    \caption{ERS vs. CRS model, simulated by MiStImm 500--500 times at various initial number of pathogens. The unit of time is one-tenth of a day; {\it f cells}: the initial number of foreign cells at the time instant 3000; {\it div time}: the mean waiting time between two divisions of a foreign cell; {\it ERS wins}: number of wins of the immune system against pathogens using the ERS model setting; {\it CRS wins}: number of wins of the immune system against pathogens using the CRS model setting; {\it ratio}: ERS wins divided per CRS wins; {\it p-value}: one-sided p-value of the Fisher's exact test. In every cases ERS performed significantly better than CRS.}
    \label{tab:sim_res_1}
    \begin{tabular}{|c|c|c|c|c|c|}
        \hline
        \textbf{f cells}    & \textbf{div time} & \textbf{ERS wins}     & \textbf{CRS wins} & \textbf{ratio}    & \textbf{p-value}          \\ \hline
        200 & 50 & 499 & 432 & 1.155 & 1.1E-20 \\ \hline
        250 & 50 & 497 & 361 & 1.377 & 1.85E-42 \\ \hline
        300 & 50 & 481 & 310 & 1.552 & 5.19E-45 \\ \hline
        350 & 50 & 417 & 225 & 1.853 & 4.48E-38 \\ \hline
        400 & 50 & 272 & 135 & 2.015 & 5.74E-19 \\ \hline
    \end{tabular}
\end{table}

\begin{table}[ht!]
    \centering
    \caption{ERS vs. CRS model, simulated by MiStImm 500--500 times at various mean waiting time between two divisions of a pathogens. Column labels are the same as in Table~\ref{tab:sim_res_1} but the positions of the columns ``f cells'' and ``div time'' are switched. In every cases ERS performed significantly better than CRS.}
    \label{tab:sim_res_2}
    \begin{tabular}{|c|c|c|c|c|c|}
        \hline
        \textbf{div time}    & \textbf{f cells}& \textbf{ERS wins}   & \textbf{CRS wins} &\textbf{ratio}     & \textbf{p-value}          \\ \hline
        40 & 350 & 208  & 66  & 3.152 & 1.19E-24 \\ \hline
        50 & 350 & 417 & 225 & 1.853 & 4.48E-38  \\ \hline
        60 & 350 & 473 & 320 & 1.478 & 1.16E-35  \\ \hline
        70 & 350 & 493 & 400 & 1.233 & 1.42E-24  \\ \hline
        80 & 350 & 500 & 441 & 1.134 & 2.81E-19  \\ \hline
    \end{tabular}
\end{table}

\begin{table}[ht!]
    \centering
    \caption{Contingency table of the one sided Fisher's exact test \cite{F1949} for the fourth row of Table~\ref{tab:sim_res_1}. The $p$-value appearing there was calculated by the formula $\sum_{i=417}^{500} \binom{642}{i}\binom{358}{500-i}/\binom{1000}{500} \approx 4.48E-38$. Note that the values of the hypergeometric distribution inside the sum are the probabilities of choosing 500 experiments out of 1000, containing exactly $i$ ERS wins of the given 642 total number of wins and also choosing exactly $500-i$ ERS losses of the given 358 total number of losses.}
    \label{tab:Fisher}
    \begin{tabular}{|c|c|c|c|}
        \hline
        \textbf{}    & \textbf{ERS}  & \textbf{CRS} &Row total \\ \hline
        \textbf{Win} & 417 & 225 & 642 \\ \hline
        \textbf{Loss} & 83 & 275 & 358 \\ \hline
        Col total & 500 & 500 & 1000 \\ \hline
    \end{tabular}
\end{table}


\section*{Data and code availability}
All codes and data (including the results of the simulation) are available at  \url{http://info.ilab.sztaki.hu/~kerepesi/MiStImm/}. 
Every simulation result file contain the actual parameter setting. 

\section{Conclusions} \label{sec:Conc}

First we described arguments that led us to the ERS theoretical model that emphasizes the role of self in creating, maintaining and controlling immune responses to self and nonself. Then we discussed the MiStImm in silico model that was made to investigate some important characteristics of immune development, starting from conception and ending some time after birth. Finally, results of some computer experiments were discussed. An important part of the latter was the comparison of the CRS and ERS theoretical models. We think that it is likely that evolution preferred adaptive immune systems whose basic mechanism is closer to the ERS model than to a CRS model, because ERS gives better results to overcome a critical primary infection. We hope that our ideas and our computational model may encourage investigations about the problems raised in this paper, using both in vitro and in vivo experiments. We would especially like to see experiments clarifying questions about self--nonself discrimination in a primary infection.

\section*{Acknowledgements}
We are grateful to Dr. Andrea Lad\'anyi for her useful criticism and to Prof. P\'eter Csermely for his useful advice. We thank former students G\'abor Szabados, Tam\'as Kiss, Krist\'of H\"or\"omp\"oly, and Endre Szecsei for their contributions to developing the software package. 
Csaba Kerepesi was supported by Momentum Grant of the Hungarian Academy of Sciences (LP2012-19/2012).

\section*{References}

\bibliographystyle{unsrt}
\bibliography{sample}

\end{document}